# Random Pulse Train Spectrum Calculation Unleashed

Sander Stepanov

sander.stepanov@gmail.com

Anastasios Venetsanopoulos

Ryerson University

Abstract

For the first time the problem of the full solution for the calculation of the power spectrum density of the random pulse train is solved. This well known problem led to a mistaken publication in the past and even its partial solution was considered worthy of publication in a textbook.

The little known solution for only the continues random pulse train spectrum is explained by examples and is extended to cover each signal having a discrete spectrum, too. A developed approach is used to derive the general equation for two important representative pulse trains with unbalanced symbol duration: a signal with stretched pulse with a transition from “one” to “zero”, and shortened blank symbols. The developed theoretical results are validated by simulation. It is shown that the pulse trains under consideration pose spectrum peaks. The characteristics of these peaks are investigated.

## I. INTRODUCTION

Under the condition when the digital signal is modulated by non-return to zero binary Pulse Amplitude modulation (PAM), its spectrum does not have spectrum harmonics at clock frequency, which is a significant disadvantage in the sense of timing synchronization. In order to overcome this, the following methods can be used: a continuous wave with clock frequency adding to the signal at the transmitter output; non-uniform symbol duration; pulse front predistortions during transmission; non-linear signal processing at receiver [1]. Signal formation during transmission is preferable since it permits linear signal processing at the receiver that is simpler than a non-linear one. Among these methods, the non-uniform symbol duration method is better, since it can be more easily implemented at digital logic level design (it just shortens the delay for some symbols) and for all practical purposes does not change peak to average ratio. In this investigation analytical analysis of the spectrum for specific non-uniform symbol duration PAM modulation is developed. A pulse with amplitude one is transmitted for symbol one and a blank space corresponds to symbol zero. This can be the theoretical basis for other spectrum calculation problems such as non-uniform symbol duration modulation, digital pulse interval modulation techniques [2], or radar recognition and target identification [3-4].

The suggested technique is novel and, in contrast to [5, 6], it is free from semi-empirical qualitative choices of elementary signals in the underlying model. Note that [6], which is based on [5], is dedicated to the relatively simple case, with pulse duration imbalance for only one transition from "1" to "0" without a shifting symbol time position. However even in this case, no strictly mathematical model description led to mistaken spectrum results, as when [6] corrects the results of a previous publication. When, along with this case, we consider the much more complicated pulse train with shift for all pulse time positions due to non-uniform duration of blank symbol intervals, the decision is elegant and concise. This opens the possibility to using a developed approach for other pulse train cases.

First, we consider the pulse train from [6] with unequal probability of "one" and "zero" and with pulse end shifted by value $\Delta$, at transition from "one" to "zero". Consequently, this transition creates the discrete spectrum components in overall psd. A strictly mathematical approach for this kind of pulse train is developed, since the signal consists of both periodic and non periodic parts [7-9], but the existing theory, partially explained in [10], drawing on an authoritative source [11], deals with only the non-periodic part and yields only the continuous part of the spectrum. Moreover, [11] is hardly known; therefore practical example which me give based on [11] is useful.

In the next case the symbols “one” and “zero” are equaly probable and the symbol duration predistortion under investigation supposes the pulse duration to be constant, but the blank symbol duration is cut by the small value $\Delta$. As a result, the pulse train shape is more complicated but again the spectrum of the uniform symbol duration signal is changed to yield a remarkable features. This feature, the peak, is close to clock frequency, whose amplitude and frequency shift are dependent on Δ.

In these cases the PSD peaks bear information about the clock frequency and make possible receiver linear signal processing in order to recover the clock frequency. The dependence of investigated spectrum peak parameters on $\Delta$ is considered.

The paper is organized as follows: section 2 relates to PSD calculation for unbalanced transaction from “1” to “0”; section 3 focuses on PSD calculation for unbalanced all zero symbols; and section 4 concludes the paper.

## II. PSD Calculation For Unbalanced Transaction From One to Zero

The unbalanced pulse train consists of *n* non-overlapped NRZ rectangular shape pulses $S_r(t,\tau_\upsilon)$, with duration $\tau_\upsilon$. It can be described by using the pulses on the time axis

$$S(t-t_\upsilon,\tau_\upsilon) = S_r(t,\tau_\upsilon)\,, \quad \text{for} \quad t_\upsilon \le t \le t_\upsilon + \tau_\upsilon\,;$$

$$S(t-t_\upsilon,\tau_\upsilon) = 0, \ \text{for} \quad t < t_\upsilon, \quad t > t_\upsilon + \tau_\upsilon\,; \qquad (1)$$

then this pulse train is described as

$$\xi(t) = \sum_{\upsilon=1}^{\upsilon=n} S(t-t_\upsilon,\tau_\upsilon)\,, \qquad (2)$$

with overall duration

$$T_n = \sum_{\upsilon=1}^{\upsilon=n} \vartheta_\upsilon \,, \qquad (3)$$

where $\vartheta_\upsilon$ is the distance between the fronts of pulse number $\upsilon$ and the next pulse.

In the general case, a signal without DC consists of non-periodic and periodic parts, so that the PSD consists of continuous and discrete parts [7, 8].

According to the PSD definition, the PSD of the non-periodic part of a signal is calculated by [9]

$$S_c(\omega) = \lim_{T_n \to \infty} \left\langle \frac{\left| F_{T_n}(\omega) \right|^2}{T_n} \right\rangle \,, \qquad (4)$$

where

$\langle \bullet \rangle$ stands for averaging;

and

$F_{T_n}(j\omega) = \int_{\frac{-T_n}{2}}^{\frac{T_n}{2}} \xi(t) e^{-j\omega t} dt$ - is the spectrum for pulse train realization $\xi(t)$.

Taking into account the fact that

$$F_{T_n}(j\omega) = \sum_{\upsilon=1}^{\upsilon=n} F_r(j\omega, \tau_\upsilon) e^{-j\omega t_\upsilon} \,, \qquad (5)$$

where $F_r(j\omega, \tau_\upsilon)$ is the spectrum of $S_r(t, \tau_\upsilon)$,

equation 5.42 from [10] can be used to derive the following

$$\left\langle \left| F_{T_n}(j\omega) \right|^2 \middle| \ n \right\rangle = \sum_{\mu=1}^{\mu=n} \sum_{\upsilon=1}^{\upsilon=n} \left\langle F_r(j\omega, \tau_\upsilon) F_r^*(j\omega, \tau_\mu) e^{j\omega(t_\mu - t_\upsilon)} \right\rangle =$$

$$= \left\langle F_r(j\omega, \tau_\upsilon) \right\rangle \left\langle F_r^*(j\omega, \tau_\mu) \right\rangle \sum_{\mu=1}^{\mu=n} \sum_{\upsilon=1}^{\upsilon=n} \left\langle e^{j\omega(t_\mu - t_\upsilon)} \right\rangle . \qquad (6)$$

The value of $\left|\left\langle e^{j\omega(t_\mu - t_\upsilon)}\right\rangle\right| = \left|\Theta(\omega,\vartheta)\right|^{|t_\mu - t_\upsilon|}$, where $\Theta(\omega,\vartheta)$ is the characteristic function of the distance between neighboring pulse fronts (see eq. 5.45 from [10] ) can be calculated by using the characteristic functions of the pulse duration

$$\Theta_1(\omega,\tau) = \int_0^\infty f_1(\tau) e^{jw\tau} d\tau$$

and the distance between neighboring pulses

$$\Theta_2(\omega,\lambda) = \int_0^\infty f_2(\lambda) e^{jw\lambda} d\lambda \,,$$

since $\vartheta = \tau + \lambda$.

By observing that

$$\Theta(\omega,\vartheta) = \Theta_1(\omega,\tau)\ \Theta_2(\omega,\lambda) =$$

$$qp\left(\frac{e^{j\omega T}}{1-pe^{j\omega T}}\right)\left(\frac{e^{j\omega T}}{1-qe^{j\omega T}}\right) = \Theta(\omega=\omega_0=\frac{k2\pi}{T_0},\vartheta)=1 \,, \qquad (7)$$

it can be concluded that

$$\left|\sum_{\mu=1}^{\mu=n}\sum_{\upsilon=1}^{\upsilon=n}\left(\left\langle e^{j\omega(t_\mu - t_\upsilon)}\right\rangle = 1\right)\right| = n^2 \to \infty \,, \qquad (8)$$

for $n \to \infty$ and $\omega = 2\pi / T_0$.

So the PSD of the signal under investigation can have the discrete component (previously referred to as “part”) when the first multiplier in (6) is nonzero. This leads to the conclusion that the signal can have a periodic component with period $T_0$ [7, 8].

Turning to the periodic part of the signal with period $T_0$, its discrete PSD [9] can be derived as

$$S_d(\omega\ ) = \left\langle 2\pi \sum_{n=\infty}^{n=-\infty} |F_n|^2\ \delta(\omega - n\omega_0)\right\rangle \quad , \quad (9)$$

where

$F_n$ are Fourier series coefficients;

$\omega_0 = 2\pi / T_0$.

By analogy with (6), the following equation can be written

$$\left\langle \left| F_n\ ( j\omega = j\omega_0 k = j\omega_k ) \right|^2 \ n \right\rangle = \sum_{\mu=1}^{\mu=n}\sum_{\upsilon=1}^{\upsilon=n} \left\langle \frac{F_r(j\omega_k,\tau_\upsilon)}{T_n} \frac{F_r^*(j\omega_k,\tau_\upsilon)}{T_n} e^{j\omega_k(t_\mu - t_\upsilon)} \right\rangle =$$

$$\left\langle \frac{F_r(j\omega_k,\tau_\upsilon)}{T_n} \right\rangle \left\langle \frac{F_r^*(j\omega_k,\tau_\upsilon)}{T_n} \right\rangle \left\langle \sum_{\mu=1}^{\mu=n}\sum_{\upsilon=1}^{\upsilon=n} e^{j\omega_k(t_\mu - t_\upsilon)} \right\rangle . \quad (10\ )$$

Now it is necessary to find

$$\left\langle \sum_{\mu=1}^{\mu=n}\sum_{\upsilon=1}^{\upsilon=n} e^{j\omega_k(t_\mu - t_\upsilon)} \right\rangle = \left\langle n^2 \right\rangle . \qquad (11)$$

The observation that $\left\langle n^2 \right\rangle \approx \left\langle n \right\rangle^2$ can be used to resolve the calculation of $\left\langle n^2 \right\rangle$.

Further calculation of $\left\langle n \right\rangle$ can be done, by using

$$\lim_{n\to\infty} \frac{\left\langle T_n \right\rangle}{T_n} = 1 \ , \qquad \left\langle n \right\rangle = \frac{\left\langle T_n \right\rangle}{\left\langle \vartheta \right\rangle} , \quad (12)$$

as in [10].

The first multiplier from (10) is

$$\left\langle F_r^*(j\omega_k,\tau) = \frac{j(e^{-j\omega_k\tau} - 1)}{\omega_k} \right\rangle = \frac{j}{\omega_k}\left( \left\langle e^{-j\omega_k\tau} \right\rangle - 1 \right) = \frac{j}{\omega_k}(\Theta_1^*(\omega_k,\tau) - 1) , \qquad (13)$$

where

$$\Theta_1^*(\omega_k,\tau) = \int_0^\infty f_1(\tau) e^{-j\omega_k\tau} d\tau .$$

Similarly the following can be calculated

$$\left\langle F_r\ (j\omega_k,\tau) = \frac{j(e^{-j\omega_k\tau}-1)}{\omega_k}\right\rangle = \frac{j}{\omega_k}(\Theta_1(\omega_k,\tau)-1)\ . \quad (14)$$

Eventually for the general case (15) can be derived

$$S_p(\omega=\omega_k) = \frac{2\pi\frac{j}{\omega_k}(\Theta_1(\omega_k,\tau)-1)\frac{j}{\omega_k}(\Theta_1^*(\omega_k,\tau)-1)}{\langle\vartheta\rangle^2}\delta(\omega-\omega_k)\quad . \quad (15)$$

For the current pulse train, the key part of (15) can be simplified to

$$\frac{j}{\omega_k}(\Theta_1(\omega_k,\tau)-1)\frac{j}{\omega_k}(\Theta_1^c(\omega_k,\tau)-1) = \frac{j}{\omega_0 n}(e^{-j\omega_0 k\Delta}-1)\frac{-j}{\omega_0 n}(e^{j\omega_0 k\Delta}-1) = 4\left[\sin(\frac{\omega_k\Delta}{2})\right]^2\ . \quad (16)$$

Finally by using substitutions (A.2) and (A.3) and using (A.8), (17) is derived

$$S_p\left(\omega=\omega_k=\frac{k2\pi}{T_0}\right) = \frac{2\pi\sin^2(\frac{\omega_k\Delta}{2})(p(1-p))^2}{(k\pi)^2}\delta(\omega-\omega_k)\,. \quad (17)$$

For spectrum diagrams, it is more convenient to use the frequency axis measured in Hz. To transform (17) to Hz the comparison of equations (2.3.70) from [12] and (1.102) from [9] can be used, resulting in

$$S_p\left(f=f_k=\frac{k}{T_0}\right) = \frac{S_p\left(\omega=\omega_k=\frac{k2\pi}{T_0}\right)}{2\pi} = \left(\frac{\sin\ (f_k\Delta\pi)p(1-p)}{k\pi}\right)^2\delta(f-f_k)\,. \quad (18)$$

To calculate the continuous part of the PSD, the general equation from [11] can be used

$$S_c\left[\xi(t)-\langle\xi(t)\rangle,\omega\right] = 2\omega^{-2}\left[\langle l\rangle+\langle\tau\rangle\right]^{-1}\mathrm{Re}\left(\frac{(1-\Theta_1(\omega))\quad(1-\Theta_2(\omega))}{1-\Theta_1(\omega)\ \Theta_2(\omega)}\right), \quad (19)$$

where the multiplier 4 from the original equation was changed to 2, in order to harmonize the old style of the Fourier transform expression to the modern one.

Then by substitution $\Theta_1(\omega)$ and $\Theta_2(\omega)$ as per Appendix A for the particular pulse train, (20) is derived.

$$S_c\left[\xi(t)-\langle\xi(t)\rangle,\omega\right]=2\omega^{-2}\left[\langle l\rangle+\langle\tau\rangle\right]^{-1}\mathrm{Re}\left(\frac{\left(1-qe^{jw\Delta}\left(\frac{e^{jwT}}{1-pe^{jwT}}\right)\right)\left(1-pe^{-jw\Delta}\left(\frac{e^{jwT}}{1-qe^{jwT}}\right)\right)}{1-q\left(\frac{e^{jwT}}{1-pe^{jwT}}\right)p\left(\frac{e^{jwT}}{1-qe^{jwT}}\right)}\right).(20)$$

To make similar results with previously received equation (5) from [6], the amplitude should be set to 0.25 to adjust the scale.

To use the Hz axis to demonstrate simulation results it is necessary to take into account the fact that the simulation curve at each point represents the sum power of all PSD values between neighboring frequencies (see (2.23b) from [9])

$$P_c(f_k)=\frac{S_c(\omega_k)(\omega_k-\omega_{k-1})}{2\pi}\ . \quad (21)$$

It is useful to develop a way to find periodicity in signals by qualitative analysis of some signal characteristics [[[here I relate to the Soosan's question how predict the existence of spectrum spikes for signals intuitively].]] The periodicity of the time shape of a random signal is not obvious, but the periodicity in pdf for the distance between pulse fronts is noticeable and leads to spectrum spikes at periodic frequencies when the product of the first and second multipliers in (10) is zero. For example, for usual balanced modulation when $\Delta = 0$, there is the same $\vartheta$ pdf periodicity as for the case when $\Delta \neq 0$; however, the sine argument in (16) is zero , so there are no peaks in PSD.

To illustrate this point consider the discrete pdf of $\vartheta$.

$$\int_0^\infty f(\vartheta)dt=1=f_1\delta(\vartheta-T_0)+f_2\delta(\vartheta-2T_0)+f_3\delta(\vartheta-3T_0)...,f_N\delta(\vartheta-NT_0)\ . \qquad (22)$$

Analytically it can be expressed that the characteristic function is equal to 1 ( so (8) gives the PSD spike) , in this case, for some frequency $\omega_0$ , as follows

$$\left|\Theta(\omega_0 = \frac{2\pi}{T_0})\right| = \sqrt{\int_0^\infty (f(t)\cos(\omega_0 t)dt)^2 + \int_0^\infty (f(t)\sin(\omega_0 t)dt)^2} =$$

$$= \sqrt{\sum_{i=1}^{i=N}(f_i \cos[\omega_0(t_1+(i-1)T_0)])^2 + \sum_{i=1}^{i=N}(f_i \sin[\omega_0(t_1+(i-1)T_0)])^2} = \sqrt{(\cos\omega_0 t_1(\sum_{i=1}^{i=N} f_i))^2 + (\sin\omega_0 t_1(\sum_{i=1}^{i=N} f_i))^2} =$$

$$= \sqrt{[(\cos\omega_0 t)^2 + (\sin\omega_0 t)^2]\sum_{i=1}^{i=N} f_i} = 1. \qquad (23)$$

The validation simulation was accomplished as in [6], by averaging a large number of spectra realizations, drown from signals, in which each symbol consists of 64 samples. Fig.1 and Fig. 2 depict the simulation results for the minor and the significant difference of probability “one” and “zero”. In Fig. 1 the discrete spectrum is based at local minimums of the continuous spectrum, which is not the case in Fig. 2. From these figures it is possible to see that this simulation resolution does not yield the same results as would be yielded by calculation. In order to get the convergence of simulated and calculated curves, the FFT size was doubled in Fig. 3, but the convergence was not achieved. Only by doubling the number of samples for one symbol, the theoretical and experimental results converge as shown by Fig. 4. This observation is important for forthcoming simulations, specifically for electromagnetic compatibility problems.

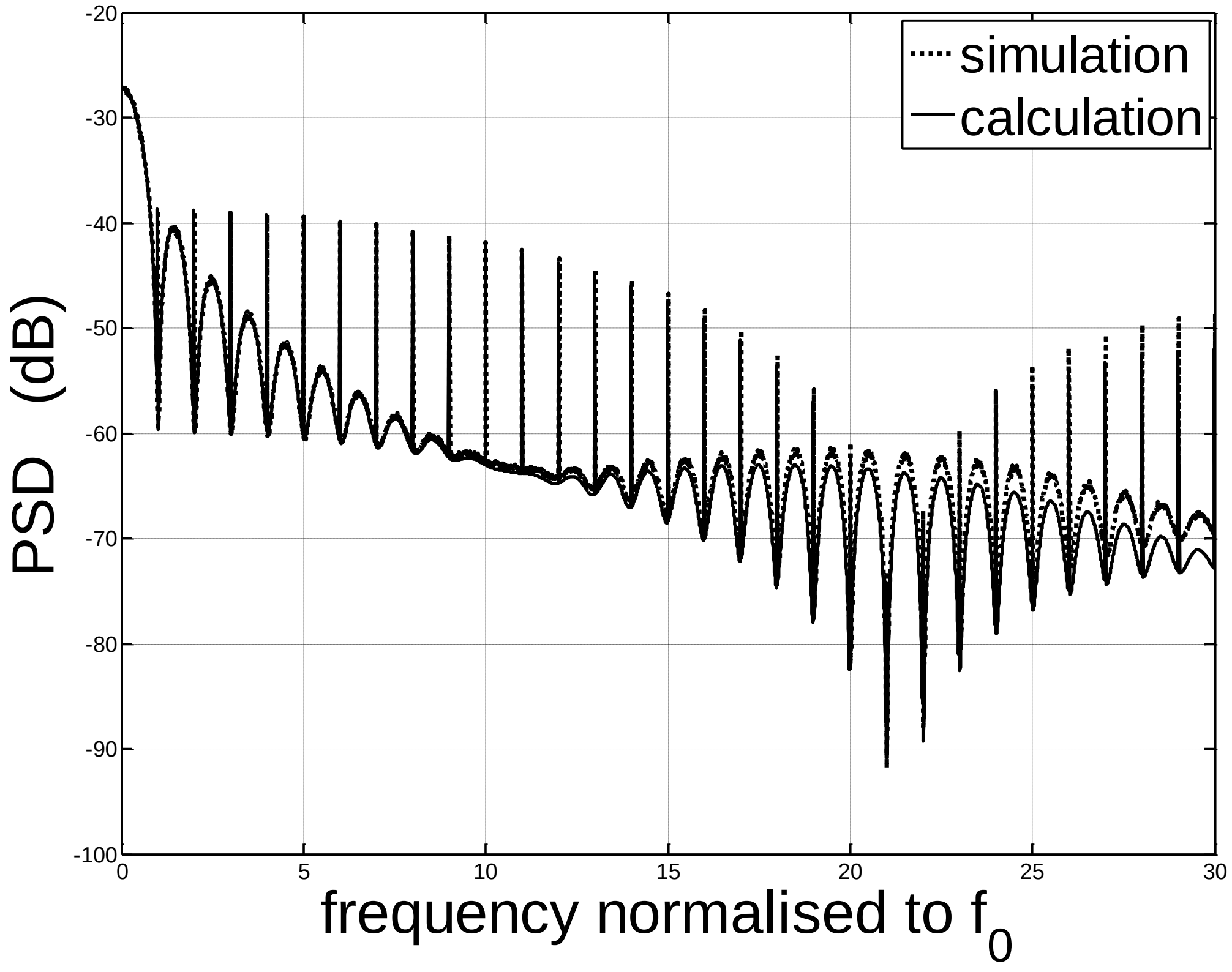

Fig. 1. Simulation and calculation results for: $T_0$ = 64, $\Delta$ = 3, $P$ = 0.55 and FFT size equal to 8192. The sampling frequency is 1 Hz.

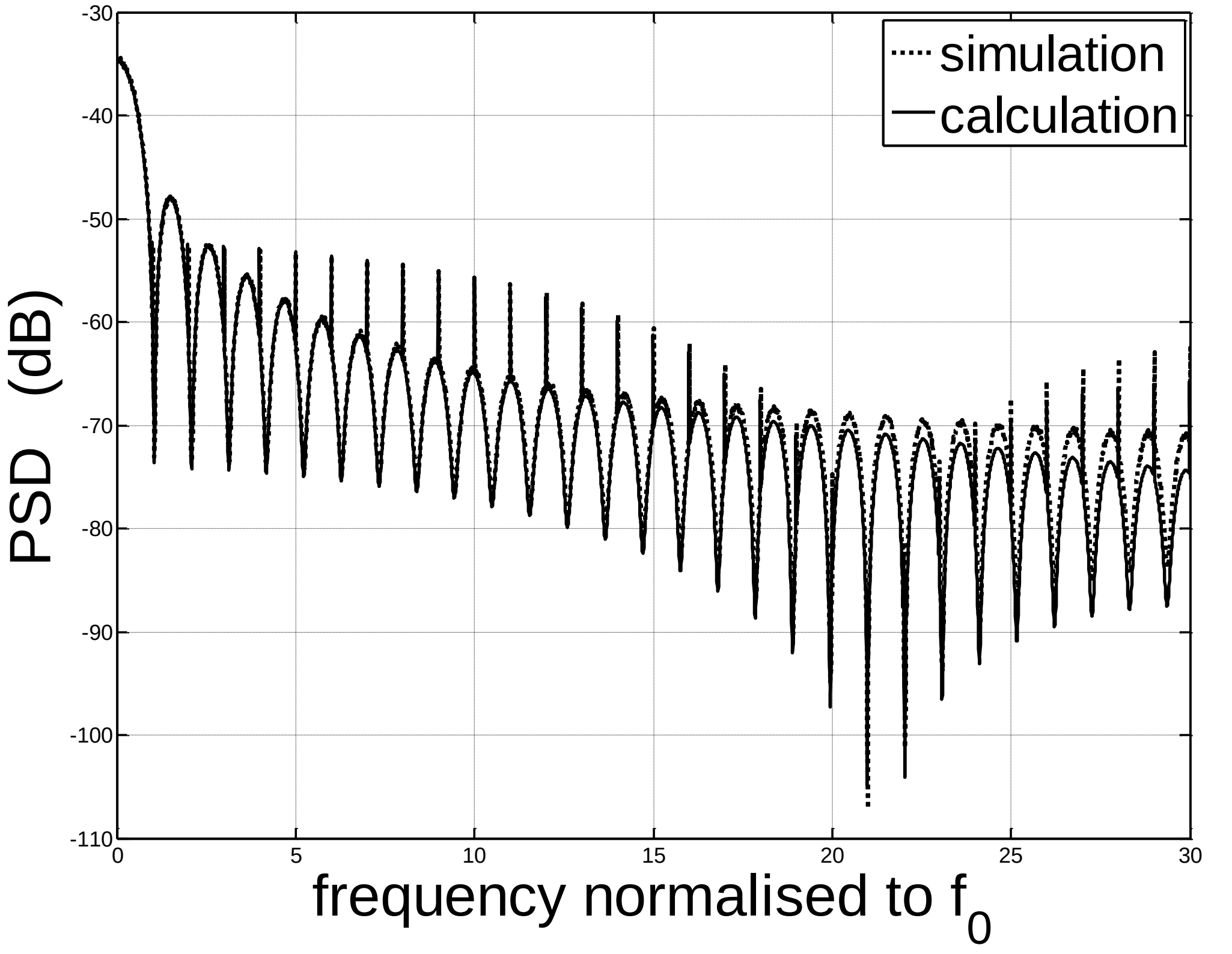

Fig. 2. Simulation and calculation results for: $T_0$ = 64, $\Delta$ = 3, $P$ = 0.95 and FFT size equal to 8192. The sampling frequency is 1 Hz.

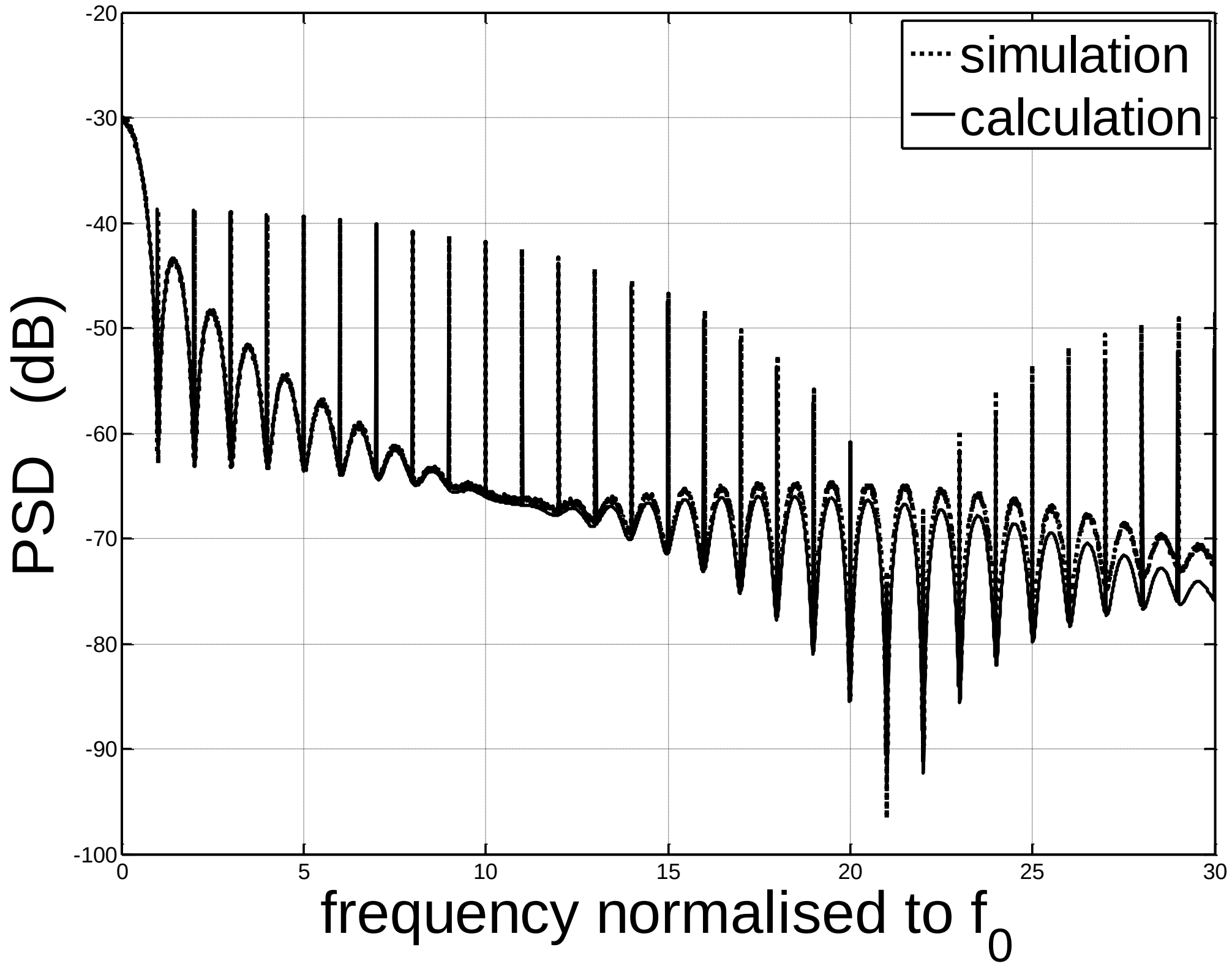

Fig. 3. Simulation and calculation results for: $T_0$ = 64, $\Delta$ = 3, $P$ = 0.55 and FFT size equal to 16384. The sampling frequency is 1 Hz.

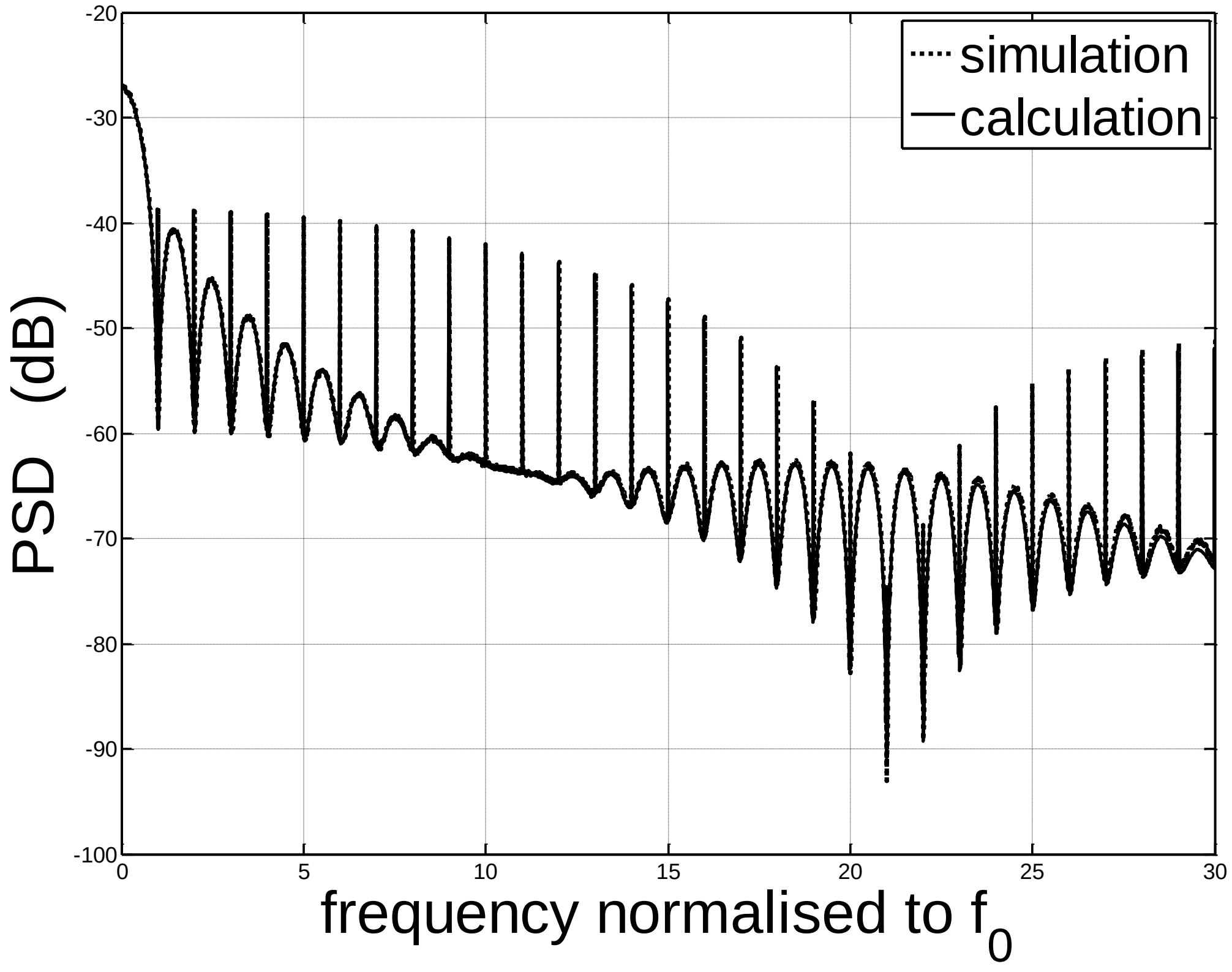

Fig. 4. Simulation and calculation results for: $T_0$= 128, $\Delta$ = 6, $P$ = 0.55 and FFT size equal to 16384. The sampling frequency is 1 Hz.

# III. PSD CALCULATION FOR THE CASE WHERE ALL ZERO SYMBOLS ARE UNBALANCED

The probability density function of intervals between pulse fronts when the “zero” symbol duration is reduced by variable Δ and “one” has the constant length $T_0$ is

$$f(l)= \sum_{k=1}^{k=\infty} \delta(l- ( k*T_0 - \Delta_k ) ) / 2^k , \qquad (22)$$

where

$\Delta_k$ represents cutting values in increasing order $\Delta_1 < \Delta_2 < \Delta_3 \ldots$ .

Thus, the characteristic function of $f(l)$ is

$$\Theta(\omega) = \sum_{k=1}^{\infty} e^{j\omega(kT_0-\Delta_k)} / 2^k . \qquad (23)$$

Taking into account that

$$\Delta_k = \Delta * k, \quad k = 2 ,3 ,\ldots ,\infty ;$$

$$\Delta_k = 0 \quad k = 1, \qquad (24)$$

the following can be calculated

$$\Theta(\omega) = \frac{e^{j\omega T_0}}{2} + \sum_{k=2}^{k=\infty} \frac{e^{j\omega(kT_0-\Delta k)}}{2^k} =$$

$$\frac{e^{j\omega T_0}}{2} - \frac{e^{j\omega(2T_0-\Delta)}}{2(e^{j\omega T_0} - 2e^{j\omega\Delta})}, \qquad (25)$$

Where the series limit was found by using (0.231) from [15].

According to [11, 14] the PSD of interest is

$$S(\omega) = K|F(j\omega)|^2 \Phi(\omega), \, (26)$$

where $K$ is a constant;

*F(j ω)* is the power spectrum density of a pulse which represents "1";

$$\Phi(\omega) = \mathrm{Re}\left(\frac{1+\Theta(\omega)}{1-\Theta(\omega)}\right).$$

Finally, by substiting (25) into (26) for rectangular pulses

$$S(\omega) = K\left(\frac{1}{\omega}\right)\sin^2\left(\frac{\omega T_0}{2}\right) \ \mathrm{Re} \left( \frac{1+\dfrac{e^{j\omega T_0}}{2} - \dfrac{e^{j\omega(2T_0-\Delta)}}{2(e^{j\omega T_0} - 2e^{j\omega\Delta})}}{1-\dfrac{e^{j\omega T_0}}{2} + \dfrac{e^{j\omega(2T_0-\Delta)}}{2(e^{j\omega T_0} - 2e^{j\omega\Delta})}} \right). \qquad (27)$$

In order to validate the equation (27), the spectrum of the signal under investigation was determined by simulation. The simulation parameters are: signal length is 2000 symbols; the spectrum is calculated by averaging spectra from 500 signal realizations; each spectrum is determined through a correlation function in order to synchronize the spectrum axes of different lengths of signal realization; symbol duration is 100 seconds; sampling rate is 1 second.

As it is possible to see from Fig. 1, the analytical and simulated spectra are similar.

By comparing signal spectra for original and predistorted cases, it follows that the micro-level parameter — small changes of blank symbol duration — leads to the existence of a macro parameter – frequency peak at clock frequency — which can be used as a clock reference in the receiver.

For analytical investigation the following next peak parameters were used: amplitude, width, and central frequency. Peak amplitude is approximately twice as high as the second lobe maximum value (Fig. 1). From Fig. 2 it can be seen that the peak amplitude increases significantly when $\Delta$ is decreased. As is shown in Fig. 3, the peak width is significantly narrowed when $\Delta$ is decreased. The dependence of the normalized central frequencies of the main peaks on the values of $\Delta$ is shown on Fig. 4. It is possible to see that these frequencies are almost linearly dependent on $\Delta$.

To sum up, from a symbol timing point of view, a trade off optimization should be carried out to find the most suitable value of $\Delta$ for each specific requirement and capability of the symbol synchronization system.

The additional potential fields of application for the suggested theoretical result can be used in different areas where the random pulse train is used or where it can be brought in for use with the spectrum peaks feature. For example, three potential fields might be described as follows:

[[[ further three paragraphs are more science fiction than science, but here I would prefer to write as general as possible, for example the term "radar" can stand for usual electromagnetic waves radar , sonar systems and for ultrasound ( not only medical ultrasound) , too. The "killer application" for this material can be in unexpectable area. Really , I think we will not put this three paragraphs in paper, but I spent a lot of time to develop them, since it can play important role . Mostly I am suspicious in pulse symbol duration changing due to radar or/and target moving, but who knows… ]]]

1. When the receiver and transmitter are moved in space (radar field [15] or wireless data transmitted from a moving object or between moving objects), this moment causes a change in the distance between them, and as a result, a change in symbol duration. Then parameters of speed can be evaluated by peak analysis, i.e. the data signal can be used as an indicator signal for velocity and its own characteristic estimation. Moreover, in the case when the velocity changes, it is not enough to make needed symbol duration changes, the information about speed can be especially inserted by modulation of blank symbol duration by a transmitter. In another way at the receiver side, this information can be extracted from the signal when the Doppler frequency shift is transformed to pulse duration spread. (A nonlinear phase characteristic can be applied.)

2. For target tracking in a radar field, the random data transmitted by the target (reflected or not reflected) can be used as a radar signal to hide the fact of tracking. In other words, the random pulse train can be used as a radar signal as an electronic countermeasure in jamming conditions [16, 17], in order to change the deterministic radar signal nature to a random radar signal basis. This is done in order to add some chaos to the signal to makes it more resistant to random jamming, since the random nature of the radar signal with stands in counterposition to the random nature of jamming jamming.

3. For data signal propagation, the identification of the environmental parameters for nonlinear dispersive channels [18, 19 (maybe 20, too)] is the next potential field. For this kind of environment the blank pulse shortening value will depend on the dispersion of nonlinearity values, since the pulse spreading depends on these parameters. As a result, by using the spectrum peak analysis, the dispersion of the communication channel can be estimated without using additional signals for channel testing. This special spreading can be created by pulse spreading due to random pulse train modulation by some parameter, which causes different pulse spreading depending on the changes in simultaneous time value of this parameter. For example, slight tooth (triangular) modulation in time with period significantly greater than symbol duration will generate a linearly increasing/decreasing amplitude. As a consequence, it will originate blank pulse shortening, where pulses with different amplitude will be stretched differently. As result the distance between pulses becomes bigger/smaller as the pulse spreading becames bigger/smaller.

Here in pictures I use the numbers starting form 1, since we do not decided about above pictures

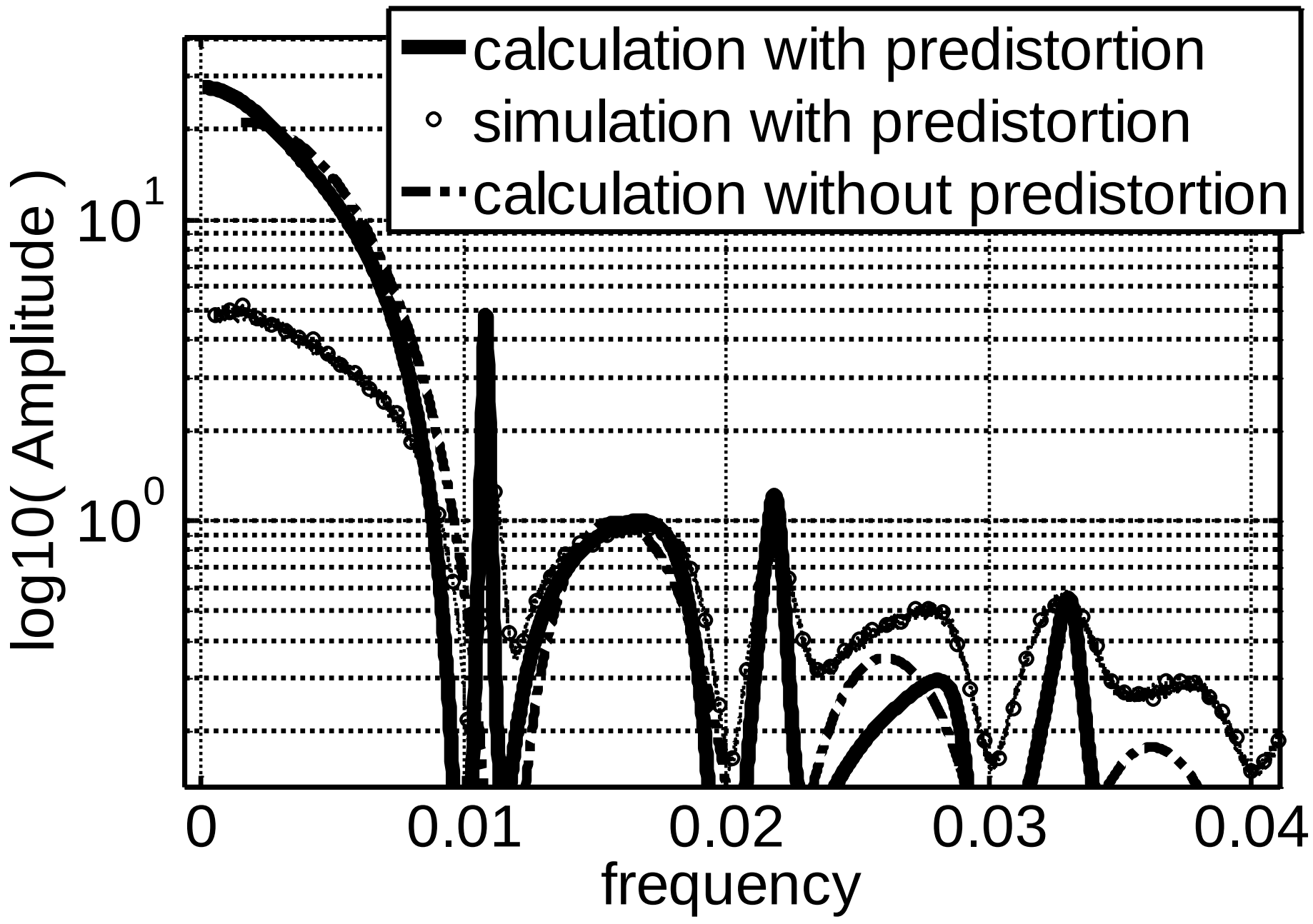

Fig.1. Calculated and simulated spectra with predistortion and a calculated spectrum for uniform symbol duration. All spectra are normalized in the following way: the second lobe amplitude is normalized to value 1. Simulation using 2000 symbols, 500 times and averaging; $T_0$ = 100 seconds; Δ = 10 seconds, rectangular pulses.

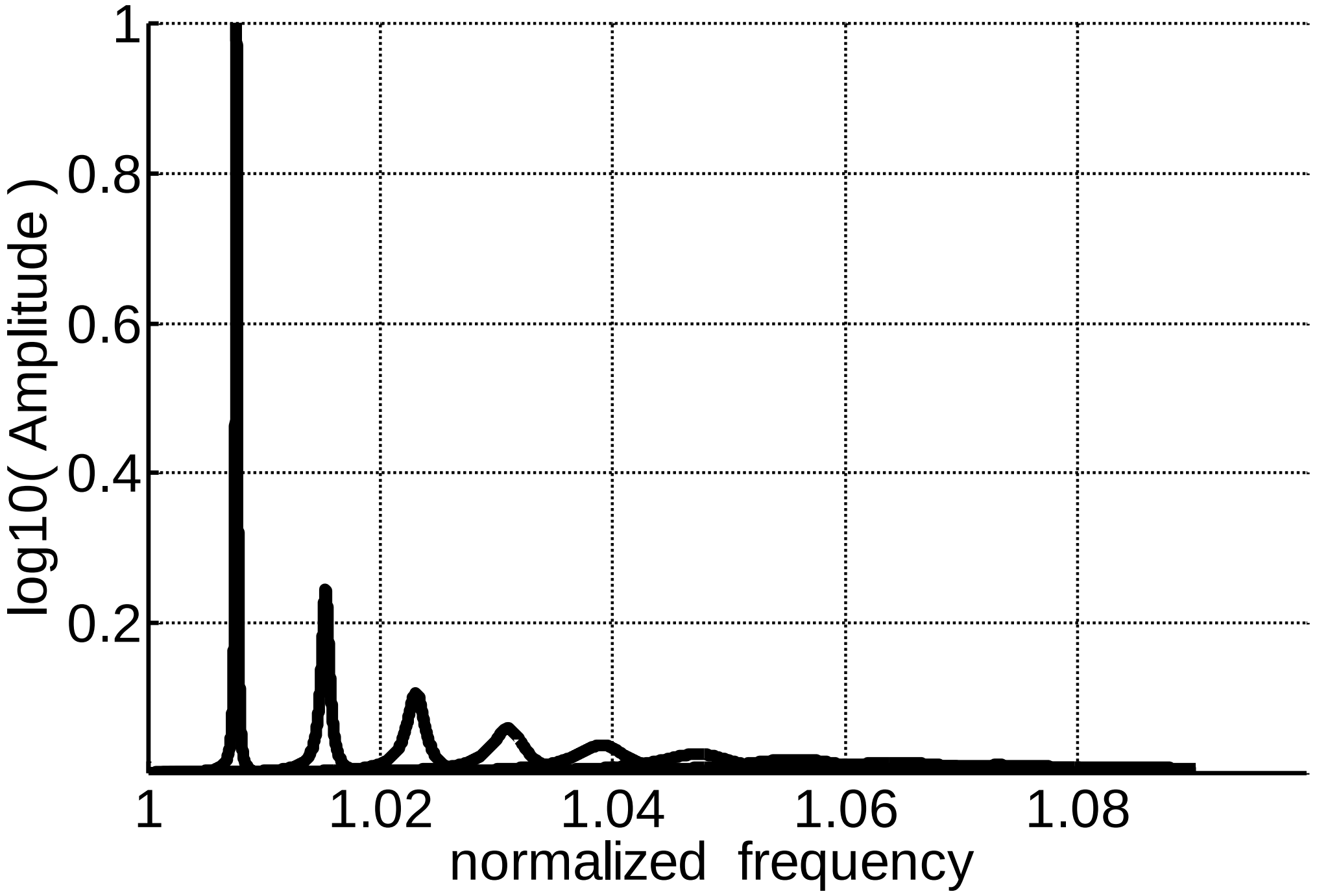

Fig. 2. Normalized peak amplitude dependence on Δ for rectangular pulses.

From left to right Δ changes from 1% to 10% of $T_0$.

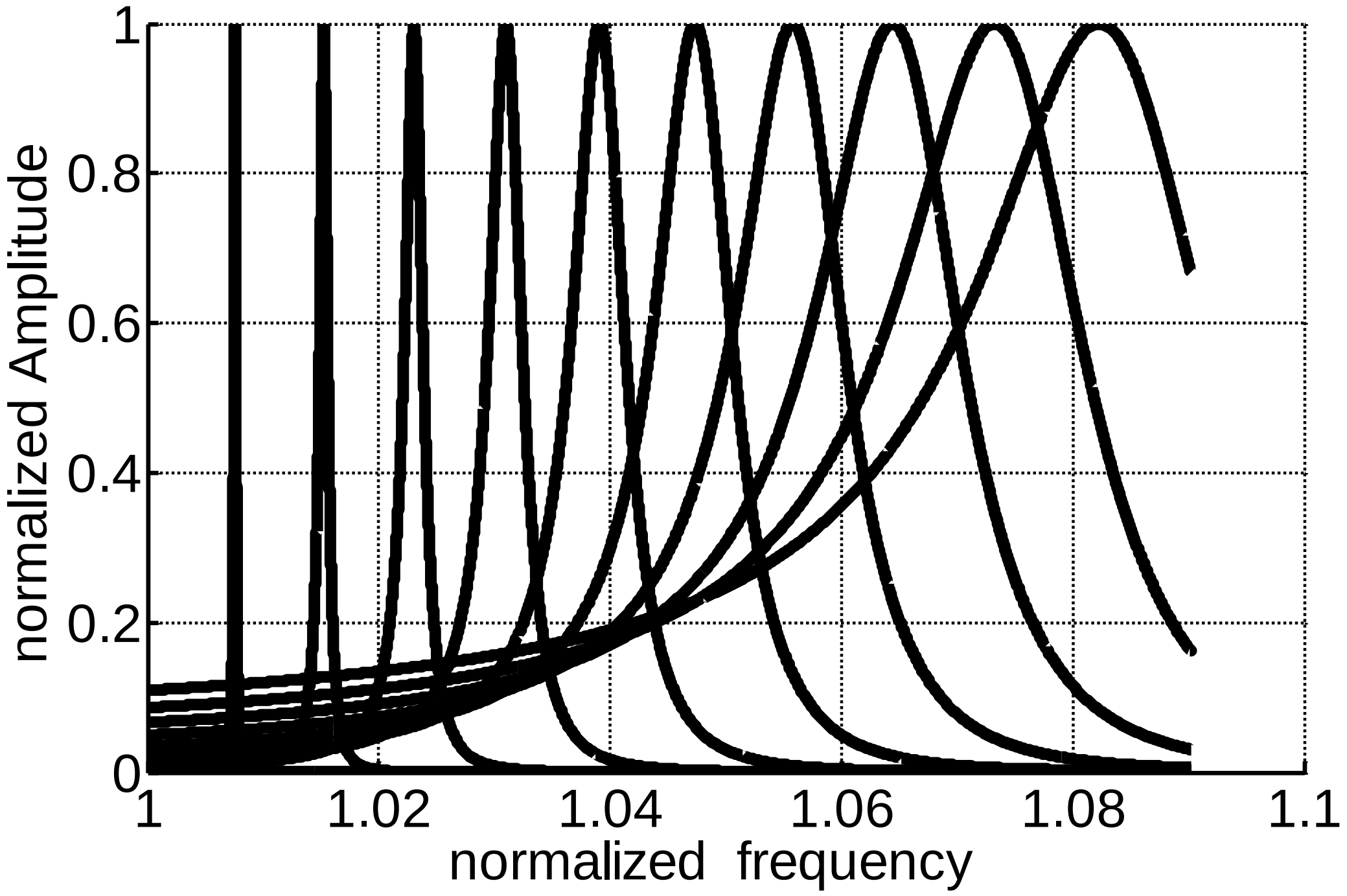

Fig. 3. Peak forms normalized to maximum values, showing dependency on Δ.

From left to right Δ changes from 1% to 10% of $T_0$ .

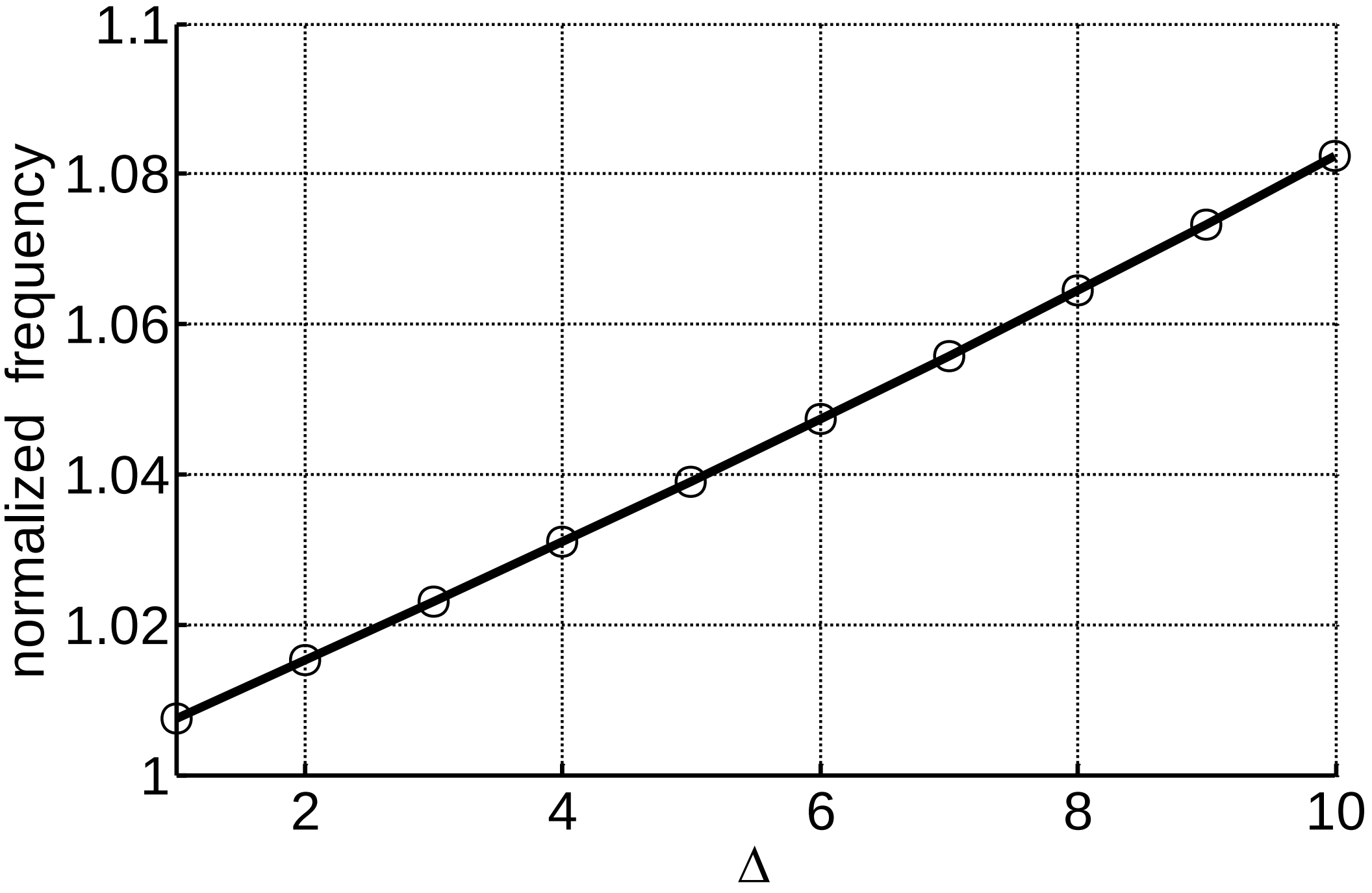

Fig. 4

The dependence of main peak normalized central frequency on the value of Δ (rectangular pulses).

## V. CONCLUSION version 1

At first we described the full decision of how to calculate the power spectrum density of the random pulses train. The approach to the spectrum calculation of a PAM signal with a shortened blank symbol, for a signal with periodicity and without, is developed. The suggested material has made possible a analytical analysis of any random pulse train and symbol timing synchronization design by using a spectrum clock peak feature. Moreover, for unbalanced all zero symbols in definite communication system design, the swapping of Δ can be implemented to balance the clock sliding due to permanent symbol duration shortening. This swapping can be used as sub-modulation to carry additional information coded in Δ, i.e. in the spectrum's peak shift sign relative to clock frequency, for example, for frame synchronization.

A general decision for problem of calculating of random pulse train spectrum is suggested. This problem is important since the random pulse trains are used to bear the information, when the parameter of symbol duration can be subject to wanted (desirable) and unwanted (undesirable) modulation. This problem has been famous for a long time; however, only a partial decision is known.

Once the pressing and important problem of a random pulse train spectrum calculation is solved then its solution can be used as the theoretical basis for other pulse trains - the constraints and extensions will manifest them selves in accordance with practicle needs. For example, the constraint on pulse independence can be applied, or pulse amplitude behavior can be extended to have a random amplitude.

The importance of developed material is not only theoretical but is also practical, yielding S new dimensions of communication and signal processing system design. Using this new theoretical tool it is possible to bring new features to the field of radar applications and to the estimation of the parameters of the dispersive communication channels.

---

As is expected of any new theoretical result, the described approach to spectrum analysis opens new horizons not only in the obvious application area -- linear signal propagation in Communication Systems__, but also for radar application and signal transmition in nonlinear dispersive environments.

# Appendix

To calculate the Characteristic Function of pulse durations $\Theta_1(\omega)$, it necessary first to find the pdf of pulse durations τ

$$f_1(\tau) = q\sum_{i=1}^{i=\infty} p^{(i-1)}\delta(\tau - (iT_0 + \Delta)). \quad \text{(A. 1)}$$

Then according to the definition of the Characteristic Function

$$\Theta_1(\omega) = \frac{q}{p} e^{iw\Delta} \sum_{i=1}^{i=\infty} p^i (e^{jwT_0})^i = qe^{jw\Delta}\left(\frac{e^{jwT_0}}{1 - pe^{jwT_0}}\right), \quad \text{(A. 2)}$$

is found by using equation (0.231) from [15]

It is useful to find the function

$$\Theta_1^*(\omega) = \int_0^\infty f_1(\tau)e^{-jw\tau}d\tau = \frac{q}{p} e^{-iw\Delta} \sum_{i=1}^{i=\infty} \frac{p^i}{(e^{jwT_0})^i} = qe^{-jw\Delta}\left(\frac{e^{-jwT_0}}{1 - pe^{-jwT_0}}\right). \quad \text{(A. 3)}$$

For distances between pulses *l* the pdf is

$$f_2(l) = p\sum_{i=1}^{i=\infty} q^{(i-1)}\delta(l - (iT_0 - \Delta)). \quad \text{(A. 4)}$$

Then the Characteristic Function of distances between pulses is

$$\Theta_2(\omega) = \int_0^\infty f_2(l)e^{jwl}dl = \frac{p}{q} e^{-jw\Delta} \sum_{i=1}^{i=\infty} q^i (e^{jwT_0})^i = pe^{-jw\Delta}\left(\frac{e^{jwT_0}}{1 - qe^{jwT_0}}\right). \quad \text{(A.5)}$$

In order to find the mean value of distances between pulse fronts $\vartheta$, it necessary to find the mean value of pulse durations and distances between pulses

$$\langle\tau\rangle = \sum_{i=1}^{i=\infty} q(iT_0+\Delta)p^{i-1} = q\left[\sum_{i=1}^{\infty}\Delta p^{i-1} + \sum_{k=0}^{\infty}(T_0+kT_0)p^k\right] = \\ = q\left[\frac{T_0}{1-p} + \frac{T_0 p}{(1-p)^2} + \frac{\Delta}{1-p}\right] = T_0 + \Delta + \frac{pT_0}{1-p} \quad , \text{(A.6)}$$

and by analogy

$$\langle l\rangle = p\left[\frac{T_0}{1-q} + \frac{T_0 q}{(1-q)^2} - \frac{\Delta}{1-q}\right] = T_0 - \Delta + \frac{qT_0}{1-q} \quad . \quad \text{(A.7)}$$

Then

$$\langle\vartheta\rangle = \langle l\rangle + \langle\tau\rangle = T_0(2+\frac{p}{q}+\frac{q}{p}) = T_0 p(1-p) \quad . \quad \text{(A.8)}$$

Code

```
%HISTORY
%Aug_2_simul_theor_Sandy.m
%Aug_1_simul_Sandy_vmeste.m
%Aug_1_simul_Sandy_July_18.m
%JULY_18_Sandy_simul_June_8_fft__N1.m
%Sandy_simul_June_8_June_7__Proakiz_corr__N1.m
%Sandy_simul_June_8_June_7__fft__N2.m
%Sandy_simul_May_24_June_7__fft__N2.m
%Sandy_simul_May_24_June_7_good_fft.m
%Sandy_simul_May_24_May_9_good_fft.m
%Sandy_simul_Apr_30_May_9_good_fft.m
%Sandy_simul_Apr_30_Apr_23_good_fft.m
%Sandy_Apr_20_Apr_23_good_fft_simulation.m
%spectrum = spectrum + spectrum_new.^2;
%Sandy_Apr_20_Apr_13_good_fft_simulation.m
%Sandy_Apr_20_Apr_13_N1_fft_simulation.m
%Sandy_Apr_12_Apr_13_N1_test_corr_simulation.m
%Sandy_Apr_10_N2_corr_simulation_spectrum_nonuniform.m
%Sandy_Apr_10_N1_corr_simulation_spectrum_nonuniform.m
% p not equal 0.5 added
%Sandy_apr_9_N2_corr_simulation_spectrum_nonuniform.m
clear all; close all; clc;   echo off
set(0,'defaultaxesfontsize',15,'defaultaxeslinewidth',1.7,...
        'defaultlinelinewidth',2.8,'defaultpatchlinewidth',0.4...
        )
%  PARAMETERS
     num_signals =  1000 %1000  %500

     FFT_size = 8192*2 %*2^4    %8192;  %*2^4

     length_coef = 1;  %5;

    p_values = [ 0.35 0.55 0.75  0.95   ];

    my_coef = 1  %1;  %2  %NUMBER POINT IN SYMBOL multiplyer

    symbol_length = 64*my_coef;

   delta =3*my_coef  * 2^0;
 %%%%%%%%%%%%%%%%%%%%%%

    for delta_ind=1:4
```

```
    %num_signals = 1000%1000 %1300  %1000 %1000%1000

    without_main_pulses = 0  %1
    dt = 1 ; % sec
    A_tone = 0.03;
    omega_tone = 2 * pi * 0.15;
%p = 0.9  %0.55  %0.01  %0.55   %0.9 %0.55  %0.01  %  0.55 %0.01 %0.55
p = p_values(delta_ind)

%FFT_size = 8192*1*2^3   %8192*2^4*2 *2 %8192*2^10  %8192*2^0  %8192*2^8



            %^^^^^^^^^^^^^^^^^^^^^^^^^^^^^^^^^^^
            %Elapsed time is 82.328497 secon
            %         for
            %________________________
            % num_signals = 1000%1000
            % FFT_size = 8192*2^4
            % my_coef = 2^0  %2^4 %2^0
            % symbol_length = 64*my_coef
            % delta = 3*my_coef
            %________________________
            %^^^^^^^^^^^^^^^^^^^^^^^^^^^^^^^^^^^

N_symbols =   FFT_size / symbol_length  %2^13%8192/64  %2^20%13  %   8  %^8   %^10
%2^13%13 %2^20;  %      2^20=    1048576     2^15=   32768

%   num_signals = 1000*1;   N_symbols =  2^10;         Elapsed time is 36.589886 seconds.

%num_signals = 1000*1;      N_symbols =  2^10 * 2^3 =  8192;   Elapsed time is 313.313885 seconds.
if    without_main_pulses
    symbol_one(1:symbol_length) = 0;

else
  symbol_one(1:symbol_length) = 1;
end
symbol_zero(1:symbol_length) = 0;
short_symbol_zero(1:symbol_length) = 0;


short_symbol_zero(1:delta) = 1;

sig_length =  N_symbols*symbol_length;
signal(1:sig_length) = -1; % array  intitializaion
```

```
%spectrum(1:N_symbols*symbol_length*2-1) = 0 + 0i; % array  intitializaion
spectrum(1:N_symbols*symbol_length) = 0 + 0i; % array  intitializaion
%my_corr  (1:N_symbols*symbol_length*2-1)=0; % array  intitializaion
%corr_new(1:N_symbols*symbol_length) = 0 ; % array  intitializaion
%spectrum_new(1:N_symbols*symbol_length) = 0 ; % array  intitializaion
%----------------------------------------------------------------------
%---        simulation
tic
for  j = 1 : num_signals
    h_wait = waitbar(j/num_signals);
data =   0  > (  rand(1, N_symbols  ) - p );
previouse_symbol = 0;
current_start_sample = 1;
for ind = 1: (N_symbols )
   current_symbol = data(ind);
   if(  (current_symbol == 0)  && (previouse_symbol==1) )
    signal(current_start_sample : (current_start_sample+symbol_length -1)  ) = short_symbol_zero;
   else
       if   (current_symbol == 0)
       signal(current_start_sample : (    current_start_sample+symbol_length -1 )  ) = symbol_zero;
       end
       if   (current_symbol == 1)
       signal(current_start_sample : (current_start_sample+symbol_length -1)  ) = symbol_one;
       end
   end
    current_start_sample = current_start_sample + symbol_length;
    previouse_symbol = current_symbol;
end

%corr_new = xcorr(signal);
%my_corr = my_corr + corr_new;

% spectrum_new = fft(corr_new);
% spectrum = spectrum + spectrum_new;

% June 7  signal = signal - mean(signal);  % !!!!
%sig_length = length(signal);
% for ind = 1: sig_length
%     signal(ind) = signal(ind)    +  A_tone *  cos ( omega_tone * ind ) ;
%     %signal(ind) =   A_tone *  cos ( omega_tone * ind ) ;
% end
signal = signal - mean(signal);  % !!!!
spectrum = spectrum + abs(    fft(signal) / sig_length     ).^2;
% figure
%  plot(    spectrum    ) % jUne 7 test
```

```
end
toc
close(h_wait)
spectrum = spectrum./num_signals;



figure
if(1)
   xx= length(signal);
   f1 = 1/xx;

 h =plot(   (1:round(xx/2)).*f1./  (1/symbol_length),
10*log10((spectrum(1:round(xx/2)))),'k:','LineWidth',3)

 %%%%%%%%%%%%%%%


 % h =plot(     10*log10((spectrum(1:round(xx/2)))),'kp','LineWidth',3)

%h =plot(  (1:round(xx/2)).*f1./  (1/64),    10*log10((spectrum(1:round(xx/2)))),'kp','LineWidth',3)

  %hold on
  %plot(   (1:xx).*f1 ,    10*log10((spectrum)),'d')

  %ylabel('  signal = signal - mean(signal);   spectrum + = abs(fft(signal));')
hy=ylabel(' spectrum = spectrum + abs(    fft(signal) / signal_length     ).^2;  spectrum =
spectrum./num_signals;  10*log10((spectrum(1:round(xx/2))) ','FontSize',15)
%h= xlabel(['(FRREQUENCY -1/64) Hz , delta=  ', num2str(delta),'  Nsymbols=
',num2str(N_symbols),'     number of avaragings= ', num2str(num_signals),'   symbol length= ',
num2str(symbol_length),'  P=  ', num2str(p)],'FontSize',25)

SS= strvcat(['without main pulses=',num2str(without_main_pulses),', (FRREQUENCY/ 1/T_0) Hz ,
delta=  ', num2str(delta),'  Nsymbols= ',num2str(N_symbols)],['N of Averages = ',
num2str(num_signals),'   symbol length= ', num2str(symbol_length),'  P=  ', num2str(p),'
A=',num2str(A_tone),'FFT size=',num2str(FFT_size)])
title(SS,'FontSize',25)


 %title(['SIMULATION   f1 =   ',num2str(f1),'  Hz    FFT size=',num2str(FFT_size)],'FontSize',25)
 %saveas(gcf,'my_fig_fft_PSD.fig')
saveas(gcf,'my_fig_fft_PSD_jpg.jpg')
%picture for paper
 figure(delta_ind+10)
 h =plot(   (1:round(xx/2)).*f1./  (1/symbol_length),
10*log10((spectrum(1:round(xx/2)))),'k:','LineWidth',3)
```

```
%%%%%%%%%%%%%%%%%%%%%%


end
grid


otvet(delta_ind,1:length(spectrum(  1:round(xx/2/length_coef)  )   ))=10*log10(   (
spectrum(1:round(xx/2/length_coef))     )     );

    end  %delta_ind
    my_x = (1:round(xx/2/length_coef)).*f1./  (1/symbol_length);
    my_y = p_values;
   figure(123)
    set(0,'defaultaxesfontsize',15,'defaultaxeslinewidth',1.7,...
        'defaultlinelinewidth',2.8,'defaultpatchlinewidth',0.4...
        )
 axes_handle = waterfall(my_x, my_y,otvet)
 %colormap(gray)
 colormap([0 0 0])
 %For example, [0 0 0] is black, [1 1 1] is white,
 set(axes_handle,'LineWidth',3)
 title(['SIMULATION    \Delta = ',num2str(delta)],'FontSize',30)
xlabel('FREQUENCY Hz  /  1/T_0')
xlabel('FREQUENCY   NORMALIZED TO f_0')
ylabel('P -  PROBABILTY  OF “ONE”   ')
 %axis([0  6000  1 4   -20    0])
 %set(axes_handle,'zLim',[0 -20])

 %%%%%%%%%%%%%%%%%%%%%%%%%%
    %      THEORETICAL



echo off
set(0,'defaultaxesfontsize',15,'defaultaxeslinewidth',1.7,...
        'defaultlinelinewidth',2.8,'defaultpatchlinewidth',0.4...
        )

    for delta_ind=1:4
p=  p_values(delta_ind)  %0.35  %0.55 %0.55  %0.5 %0.55;
q = 1-p;
```

```
tau = symbol_length;
f1 = 1/FFT_size;


start_point = 1;
%finish_point = FFT_size /2
finish_point =  2000 * 100  / 100 *2 %10  %/ 10  %1%10    % 0.195*3;

%    f_step = 0.00000001;   %0.000000013;
w_step =   (2*pi)   /   tau  /  100%1000   %10000     %100000      % 0.0000001  %0.0001  %0.00001
%0.00001;   %0.000000013;
%----------------------
%simulation
figure(delta_ind)
hold on

%%%%%%%%%%%%%%%%%%%%%%%%%%%%%%%%%%%%%%%%%%%
%                  CONTINUES SPECTRUM

arrays_length = ( w_step * finish_point - w_step  * start_point) / w_step;



f_number = 0;
%for f  = f_step:f_step:0.03% 0.1 %1
  temp_1 (1:arrays_length) =0;
    temp_2(1:arrays_length) =0;
    TETA_1(1:arrays_length) =0;
    TETA_2(1:arrays_length) =0;
    numerator(1:arrays_length) =0;
    denumerator(1:arrays_length) =0;
    spectrum1(1:arrays_length) =0;
    spectrum2(1:arrays_length) =0;
    f_axis(1:arrays_length) =0;
    w_axis(1:arrays_length) =0;
    %for f  = 0.029:f_step:0.031% 0.1 %1
    %for f  = 0.0099:f_step:0.0101% 0.1 %1
   for w  = w_step  *  start_point : w_step : w_step* finish_point% 0.1 %1
       h_wait = waitbar(  w  /   (  w_step* finish_point   )  );
f_number = f_number+1;
  f_axis(f_number) =w  /  (2*pi);
  w_axis(f_number) =w  ;
    omega  = w;
    %omega =    4*  (2*pi)   /   tau

temp_1(f_number) =  exp(i*omega*tau) /  (   1  - p*exp(i*omega*tau)   );
```

```
temp_2(f_number) = exp(i*omega*tau) /  (   1 -   q*exp(i*omega*tau)   );

TETA_1(f_number) = q *  exp( i*delta * omega)   * temp_1(f_number)  ;

TETA_2(f_number) = p *  exp( -i*delta * omega) *  temp_2(f_number)   ;

numerator(f_number) = (1 - TETA_1(f_number))    *       (1  -  TETA_2(f_number));
denumerator(f_number) = 1 - (   TETA_1(f_number)   *     TETA_2(f_number)    );
spectrum1(f_number) = real(     numerator(f_number) /  denumerator(f_number)      )   /     (omega^2)
;
%spectrum1(f_number) = abs(numerator(f_number) /  denumerator(f_number)) / omega^2;
% spectrum2(f_number) = 1 / real(...
%                           1/(1 - TETA_1(f_number))+...
%                           1/(1 - TETA_2(f_number))...
%                           -1 ...
%                           )...
%                   / omega^2;



spectrum2(f_number) =  real(  (...
                          1/(1 - TETA_1(f_number))+...
                          1/(1 - TETA_2(f_number))...
                          -1 ...
                          )^-1  )...
                  / omega^2;



   end
close(h_wait)

rigth_samples1 = find((      spectrum2  < spectrum2(1))    );
spect_3 = spectrum2(rigth_samples1);
f_axis_3 = f_axis(rigth_samples1);

rigth_samples2 = find(      spect_3  > 0    );
%spect_3 ( rigth_samples2 ) = 4*spect_3 ( rigth_samples2 )./(tau*(2+p/q+q/p));
spect_3 ( rigth_samples2 ) = 2*spect_3 ( rigth_samples2 )./(tau*(2+p/q+q/p)); %July 5


%plot(f_axis_3(rigth_samples2) , 10*real(log10(     (  (spect_3 ( rigth_samples2 )  ) )  )     ),'.k')
 %plot(f_axis_3(rigth_samples2) , 10*real(log10(     (  (spect_3 ( rigth_samples2 ) *f1 ) )  )     ),'.b')
f_axis_contin = f_axis_3(rigth_samples2);
spec_contin = spect_3 ( rigth_samples2 ) *f1;
%plot(f_axis_contin  /  (1/64), 10*real(log10( spec_contin)),'b')
```

```
%plot(f_axis , 10*real(log10(     (  (spectrum2) )  )     ),'.k')

 %plot(f_axis , imag  (       log10(       (spectrum2) )        ),'dr')
 grid
 %legend('real(log10(       (spectrum2)   )     )',' imag  (       log10(       (spectrum2) )        )')
% title([' number of pi  from 1 to  ', num2str(NNN)])
 xlabel('frequency Hz','FontSize',40)
 %return




for (n = 1:40)
    discret_f(n) = n/tau;
    qq = abs(f_axis_contin - discret_f(n));

freq_number = find(qq == min(qq));

qq = abs(f_axis_contin*tau - n);
freq_number = find(qq == min(qq));
%
%      Discret_spectrum1(n) = 2 * ...
%          (   1  -    cos   (    n  * (    2* pi  *  delta   / tau  )      )    )...
%          /  ((n* 2*  pi/  tau )^2);
%      Discret_spectrum1(n) = 4*Discret_spectrum1(n)./(tau*(2+p/q+q/p));
%  %------------------------------------------------------------------
%      Discret_spectrum2(n) = (1 - exp( i*delta * n  *    2* pi     / tau)   ) *  (1 - exp( -i*delta * n  *    2*
pi     / tau)   )  /  (n* 2*  pi/  tau )^2;
%
% Discret_spectrum2(n) = 4*Discret_spectrum2(n)./(tau*(2+p/q+q/p))  /  tau^2* (2*pi);
% %-------------------------------------------------------------------------

% %**********************************************************
 Discret_spectrum_Bishop(n) = 0.5^2*  (2* p *(1-p))^2   *...
    (   sin(  n * pi  * delta   / tau  ) /  ( n * pi)   )^2;
% %***********************************************************

% Discret_spectrum_t(n) = ...
%     (   1  -    cos   (    n  * (    2* pi  *  delta   / tau  ) )  )  / ((n*pi)^2);


Discret_spectrum(n)  =...
    (  p *(1-p)* sin(   pi*discret_f(n)  * delta      ) /  ( n * pi)   )^2  ;
```

```
spec_contin(freq_number) = spec_contin(freq_number) + Discret_spectrum(n);
end

  %stairs(discret_f,10 *log10(Discret_spectrum1),'g')

 %stairs(discret_f,10*log10(Discret_spectrum2),'--b')

 %plot(f_axis_contin/  (1/tau), 10*real(log10( spec_contin)),'k')
  plot(f_axis_contin/  (1/tau), 10*real(log10( spec_contin)),'r')
grid
hl =  legend('simulation', 'calculation','FontSize',28 )
set(hl,'FontSize',35)
   SS= strvcat(' 10* log10( analitical PSD) ')
  SS= strvcat(['FREQUENCY Hz  /  1/T_0,       Delta= ',num2str(delta)],['       symbol length  = ',
num2str(tau),'       P=  ', num2str(p)])
   xlabel(SS,'FontSize',40)
 ylabel(['f1 =   ',num2str(f1),'  Hz','   FFT size=  ',num2str(FFT_size)],'FontSize',25)
 xlim([0 30])
 scrsz = get(0,'ScreenSize');
% set(gcf,'Position',[scrsz(1) scrsz(2) scrsz(3)   scrsz(4)-80])
% set(gcf,'Position',[20   100   1280-100   1024 - 200])
%set(gcf,'Position',[20   100   scrsz(3)-100   scrsz(4) - 200])
set(gcf,'Position',[20   50   scrsz(3)-10   scrsz(4) - 200])
 %%%%%%%%%%%%%%%
 % picture for paper
figure(delta_ind+10)
hold on
 plot(f_axis_contin/  (1/tau), 10*real(log10( spec_contin)),'k')
   xlabel('frequency normalised to f_0','FontSize',40)
  ylabel(' PSD   (dB) ', 'FontSize',40)
 hl =  legend('simulation', 'calculation','FontSize',28 )
set(hl,'FontSize',35)
xlim([0 30])
scrsz = get(0,'ScreenSize');
%   [left, bottom, width, height]
set(gcf,'Position',[scrsz(1) scrsz(2) scrsz(3)   scrsz(4)-80])

%set(gcf,'Position',[scrsz(1) scrsz(2)-100 scrsz(3)-190   crsz(4)-80])
%%%%%%%%%%%%%%%%%%%%%
 xx= length(spec_contin);


 coef1= 1;
```

```
 coef2 = 32;
otvet_t(delta_ind,1:length(spec_contin(  1:round(xx/coef1/length_coef) - coef2  )   ))=10*log10(   (
spec_contin(1:round(xx/coef1/length_coef) - coef2    )     )     );

 length(spec_contin(  1:round(xx/coef1/length_coef) - coef2) )

    end  % for one value of P

% my_x = (  1:round(xx/coef1/length_coef) - coef2   ).*f1./  (1/tau);
  my_x = (  1:round(xx/coef1/length_coef) - coef2   ).*f1./  (1/tau);
   my_x = (    f_axis_contin(  1:round(xx/coef1/length_coef) - coef2  )/  (1/tau)     );
 my_y = p_values;

     figure(123)
     hold on
    set(0,'defaultaxesfontsize',15,'defaultaxeslinewidth',1.7,...
        'defaultlinelinewidth',2.8,'defaultpatchlinewidth',0.4...
        )
     coef3 = 5;
 axes_handle = waterfall(my_x(1:end-coef3), my_y(1:end),otvet_t(:,1:end-coef3))
%   xlim([0 30])
%    zlim([-80 -20])
%
 %colormap(gray)
 colormap([0 0 0])
 %For example, [0 0 0] is black, [1 1 1] is white,
 set(axes_handle,'LineWidth',3)
 title(['SIMULATION    \Delta = ',num2str(delta)],'FontSize',30)
xlabel('FREQUENCY Hz  /  1/T_0')
xlabel('FREQUENCY   NORMALIZED TO f_0')
ylabel('P -  PROBABILTY  OF “ONE”   ')
 return
```